\documentstyle[12pt]{article}
\begin{document}
%
%
\begin{center}
{\Large\bf Entanglement in the Process of Stimulated Brillouin Scattering of Light}

\vspace{2mm}

  {\bf S. V. Kuznetsov, A. V. Kyusev,  O. V. Man'ko and N. V. Tcherniega}\\
{\it P. N. Lebedev Physical Institute of the Russian Academy of
Sciences\\

Leninsky Prospect 53, Moscow 119991 Russian Federation} 

\vspace{2mm}

\begin{abstract}
The process of stimulated Brillouin scattering is described by the model of two-dimensional oscillator. The phenomenon of
entanglement which appears in the photon-phonon modes after the interaction with matter is discussed.
\end{abstract}
\end{center}
 \setcounter{page}{1} \vspace*{-19pt}
 \section*{}
 \noindent
Spontaneous Brillouin scattering which is considered to be a scattering of light at the acoustic waves caused by thermal
fluctuations was investigated in 1922. Physically stimulated Brillouin scattering (SBS) is analogous to another type of
stimulated scattering of light - stimulated Raman scattering (SRS) of light in which the laser light is scattered by the
vibrational mode of the medium. The main difference between these two types of scatterings is the different frequency shift
(till about several hundred cm$^{-1}$ for SRS and $\sim 0,01 \mbox{cm}^{-1}$ for SBS). Quantum mechanical description of SBS
can be realized with the help of the Hamiltonian quadratic in photon and phonon creation and annihilation operators in which
the electromagnetic laser wave is considered to be classical. Such Hamiltonian was used \cite{k4,k5} for SRS description.
The purpose of our work is to discuss the phenomenon of entanglement which appears in the photon--phonon modes after the
interaction with medium.

The simplest phenomenological Hamiltonian which can be used for
the description of SBS can be written as \cite{k4}
\begin{equation}\label{eq.1}
\hat H=\hbar\omega_S\hat a_s^\dagger\hat a_s +\hbar\omega_k\hat a_k^\dagger
\hat a_k +\hbar\lambda\left[e^{-i\omega_L t}\hat a_s^\dagger\hat a_k^\dagger
+e^{i\omega_L t}\hat a_k\hat a_s\right]\,,
\end{equation}
where $ \hat a_s$ and $\omega_s $ are the annihilation operator
and the frequency of the Stokes photon, $\hat a_k$ and $\omega_k$
are the annihilation operator and the frequency of the acoustic
phonon, $\omega_L$ is the laser frequency, and $\lambda$ is the
coupling constant. Laser field amplitude is involved in $\lambda$.
The laser field is considered as classical and its frequency is
determined by the condition
$
\omega_L=\omega_k+\omega_S\,.
$
The damping and depletion of the laser light wave are neglected.

If the photons and phonons of the medium were in the ground state
at the initial moment of time, then after the interaction with
laser field the quadrature dispersions take the form
 \begin{eqnarray} \label{S_1}
\sigma_{p_s^2}&=&\sigma_{p_k^2}=\sigma_{q_s^2}=\sigma_{q_k^2}=
\frac{1}{2}\cosh(2\lambda\,t),\quad \sigma_{p_s q_s}=\sigma_{p_k q_k}=0\,,
\nonumber\\
\sigma_{p_s q_k}&=&\sigma_{q_s p_k}=
-\frac{1}{2}\sinh(2\lambda\,t)\cos(\omega_L\,t)\,, \\
\sigma_{p_s p_k}&=&\sigma_{p_k p_s}=-\sigma_{q_s q_k}=-\sigma_{q_k q_s}=
\frac{1}{2}\sinh(2\lambda\,t)\sin(\omega_L\,t)\,.\nonumber
\end{eqnarray}
The dispersions of the photon and phonon quadratures have no
oscillations and are  functions of coupling constant. The
cross-variances have oscillating behaviour. One of the intermode
covariances is not equal to zero, that means that statistical
dependence of mode appeared after interaction.

 If the photons were in ground state and phonons were in the
thermodynamic equilibrium state at the initial moment of time,
then at the time moment $t$ the dispersions of the photon and
phonon quadrature components will be of the form
\begin{eqnarray} \label{S_2}
\sigma_{p_s^2}&=&\sigma_{p_k^2}=\sigma_{q_s^2}=\sigma_{q_k^2}=
\frac{1}{2}\sinh^2(\lambda\,t)\left(1+
\coth(\theta)\right)+\frac{1}{2}\,, \\
\sigma_{p_s p_k}&=&\sigma_{p_k p_s}=-\sigma_{q_k q_s}=-\sigma_{q_s q_k}=
\frac{1}{4}\sinh(2\lambda\,t)\sin(\omega_L\,t)\left(1+
\coth(\theta)\right)\,,\nonumber\\
\sigma_{p_s q_k}&=&\sigma_{p_k p_s}=-\frac{1}{4}\sinh(2\lambda\,t)
\cos(\omega_L\,t)\left(1+\coth(\theta)\right)\,, \nonumber\\
\sigma_{p_s q_s}&=&\sigma_{p_k q_k}=0\,,\quad
\theta=\frac{\hbar\omega_k}{2kT}\,.\nonumber
\end{eqnarray}
The dispersions and intermode covarianses are functions of
coupling constant and temperature of the medium in this case. If
the medium was so prepared, that the phonons were in squeezed
states at the initial moment of time, then the dispersions of
photon and phonon quadratures at the time moment $t$ will be
\begin{eqnarray} \label{S_3}
\sigma_{p_s^2}&=&\frac{1}{2\eta}\,\left[\eta\cosh^2(\lambda\,t)+
\sinh^2(\lambda\,t)
\left(\eta^2\sin^2(\omega_s\,t)+\cos^2(\omega_s\,t) \right)\right]
\\
\sigma_{p_s\,p_k}&=&\frac{1}{4\eta}\,\sinh(2\lambda\,t)
\left[\eta\sin(\omega_L\,t)+
\eta^2\sin(\omega_s\,t)\cos(\omega_k\,t)+\cos(\omega_s\,t)\sin(\omega_k\,t)
\right]  \nonumber\\
\sigma_{p_s\,q_k}&=&\frac{1}{4\eta}\,\sinh^2(\lambda\,t)\sin(2\omega_s\,t)
\left(1-\eta^2\right)\nonumber\\
\sigma_{p_s\,q_k}&=&-\frac{1}{4\eta}\,\sinh(2\lambda\,t)
\left[\eta\cos(\omega_L\,t)-
\eta^2\sin(\omega_s\,t)\sin(\omega_k\,t)+\cos(\omega_s\,t)\cos(\omega_k\,t)
\right] \nonumber\\
\sigma_{p_k^2}&=&\frac{1}{2\eta}\,\left[\eta\sinh^2(\lambda\,t)+
\cosh^2(\lambda\,t)
\left(\eta^2\cos^2(\omega_k\,t)+\sin^2(\omega_k\,t)
\right)\right] \nonumber\\
\sigma_{p_k\,q_s}&=&-\frac{1}{4\eta}\sinh(2\lambda\,t)\left[
\eta\cos(\omega_L\,t)-\sin(\omega_s\,t)\sin(\omega_k\,t)+
\eta^{2}\cos(\omega_s\,t)\cos(\omega_k\,t)
                \right] \nonumber\\
\sigma_{p_k\,q_k}&=&-\frac{1}{4\eta}
\cosh^2(\lambda\,t)\sin(2\omega_k\,t)\left[1-\eta^{2}\right] \nonumber\\
\sigma_{q_s^2}&=&\frac{1}{2\eta}\left[\eta\cosh^2(\lambda\,t)+
\sinh^2(\lambda\,t)\left(\eta^2\cos^2(\omega_s\,t)+
\sin^2(\omega_s\,t)\right)\right] \nonumber\\
    \sigma_{q_s\,q_k}&=&-\frac{1}{4\eta}\sinh(2\lambda\,t)
\left[\eta\sin(\omega_L\,t)+
\sin(\omega_s\,t)\cos(\omega_k\,t)+\eta^2\cos(\omega_s\,t)
\sin(\omega_k\,t)\right] \nonumber\\
\sigma_{q_k^2}&=&\frac{1}{2\eta}\left[\eta\sinh^2(\lambda\,t)+
\cosh^2(\lambda\,t)\left(
\eta^2\sin^2(\omega_k\,t)+\cos^2(\omega_k\,t)\right)\right]\,, \nonumber
\end{eqnarray}
where $\eta$ - is the squeezing parameter. The dispersions of
photon and phonon quadratures are functions of coupling constant
and squeezing parameter and have oscillating behaviour. The
intermode covariances are not equal to zero, so the statistical
dependence of the modes takes place in this case too. The
dependence of the intermode covariances on the squeezing
parameter, shows the possibility of controlling the properties of
statistical dependence of the modes. The statistical properties
investigations and possibility of obtaining nonclassical states in
SRS process and other types of the stimulated scatterings of light
by preparing the medium in nonclassical states were suggested
\cite{perinaj}--\cite{36}.

Entangled states are the states which are constructed as a
superposition of states each of which has the wave function
expressed as a product of wave functions depending on the
different degrees of freedom.
 Different measures of entanglement were suggested last years.
In \cite{7} the following measure of entanglement was suggested
 \begin{equation}
    \label{ent_D}
    F=\sigma_{q_s q_k}^2+\sigma_{p_s p_k}^2+\sigma_{q_s p_k}^2+
\sigma_{p_s q_k}^2\,.
 \end{equation}
We will use this measure to discuss the entanglement in the
stimulated Brillouin scattering process. In the cases (\ref{S_1}),
(\ref{S_2}), (\ref{S_3}) the measures of entanglement of the modes
after the interaction of laser field with the medium are expressed
as
\[F_1=\frac{1}{2}\, \sinh^2 \left(2 \lambda\,t \right)\,,\quad
F_2=\frac{1}{8}\, \sinh^2 \left(2\lambda\,t \right)
\left( 1+ \coth \left(\theta\right)\right)^2\,,\]
\[ F_3=\frac{1}{16}\, \sinh ^2 \left( 2 \lambda\,t \right)
\left(
\left(1+\frac{1}{\eta}\right)^2\left(\eta^2+1\right)\right)\,,\]
respectively.

In \cite{8} the measure of the entanglement was defined as the
distance between the system density matrix and the tensor product
of its partial traces over the subsystem degrees of freedom. For
Gaussian states, the measure of entanglement reads
 \begin{equation}
    \label{ent_S}
    e_G=\frac{1}{4\sqrt{\det\sigma(t)}}+\frac{1}{4\sqrt{\det\tilde{\sigma}}}
-\frac{2}{\sqrt{\det(\sigma(t)+
\tilde{\sigma})}}\,,
 \end{equation}
where we present the inverse of dispersion matrix in block form to
define the matrix $\tilde{\sigma}$, i.e.,
\[ \sigma(t)^{-1}=\left(\begin{array}{cc} B&C\\ C^T&D
\end{array}\right)\,,\quad
\tilde{\sigma}=\left(\begin{array}{cc}
\sigma_1 &0\\ 0&\sigma_2\end{array}\right)\,,\]
\[ \sigma_1^{-1}=B-C D^{-1}C^T\,,\quad
\sigma_2^{-1}=D-C^T B^{-1} C\,.\] In the cases (\ref{S_1}),
(\ref{S_2}) the measures of entanglement of photon and phonon
modes after the interaction of laser field with the medium are
\begin{eqnarray}\label{eq}
 e_{G1}&=&\frac{\cos^2(\omega_L\,t)\sinh^2(2\lambda\,t)
(3\cos^2(\omega_L\,t)\sinh^2(2\lambda\,t)+2)}{(1+
\cos^2(\omega_L\,t)\sinh^2(2\lambda\,t))(4+3\cos^2(\omega_L\,t)
\sinh^2(2\lambda\,t))}\,,\nonumber\\
 e_{G2}&=&\frac{\cos^2(\omega_L\,t)\sinh^2(2\lambda\,t)
(3\cos^2(\omega_L\,t)\sinh^2(2\lambda\,t)+2\tau(\theta))}
{(\tau(\theta)+\cos^2(\omega_L\,t)
\sinh^2(2\lambda\,t))(4\tau(\theta)+3\cos^2(\omega_L\,t)
\sinh^2(2\lambda\,t))}\,,\nonumber
\end{eqnarray}
where $\tau(\theta)=2(\coth(2\theta)+1)^{-1}$.

To conclude, we obtained the result that in SBS there exists the
phenomenon of entanglement. We evaluated the entanglement measure
by two methods; in both methods, the measure of entanglement
depends on the medium parameters (for example, temperature) and
coupling constant.

The study was partially supported by the Russian Foundation for
Basic Research under Project~No.~00-02-16516.


\begin{thebibliography}{99}

\bibitem{k4}
M.~G.~Raymer and I.~A.~Walmsley, {\it Progress in Optics}, {\bf
XXVIII}, ed. E.~Wolf, Elsevier, 1990

\bibitem{k5}
O.~V.~Man'ko, Proc. SPIE  {\bf 4069} (2000) 143--153

\bibitem{perinaj}
J.~Perina and J.~Krepelka,  J. Mod. Opt. {\bf 38} (1991)
2137--2151

\bibitem{perinova} V.~Perinova and J.~Perina,  Optica Acta {\bf 28}
(1981) 769--793

\bibitem{33}
P.~Garsia-Fernandez and Peng Zhou,  J. Mod. Opt. {\bf 41} (1994)
2259--2279

\bibitem{34}
J.~Chai and G.~Guo, Commun. Theor. Phys. {\bf 30} (1998) 513--516

\bibitem{35}
X.~Hu and F.~Nori, Phys. Rev. Lett. {\bf 79} (1997) 4605--4608


\bibitem{36}
O.~V.~Man'ko and N.~V.~Tcherniega, J. Russ. Laser Res. {\bf 22}
(2001) 201--218

\bibitem{7} A.~S.~M.~de~Castro and V.~V.~Dodonov,  J. Russ. Laser
Res.
{\bf 25}  (2002) 93--121

\bibitem{8}
V.~I.~Man'ko, G.~Marmo, E.~F.~G.~Sudarshan and F.~Zaccaria,  J.
Phys. A: Math. Gen. {\bf 35} (2002) 7137--7157


\end{thebibliography}
\end{document}